\documentstyle[epsf]{l-aa}
\newcount\fignumber\fignumber=1
\def\nfig{\global\advance\fignumber by 1}
\def\fignam#1{\xdef#1{\the\fignumber}}
\fignam{\FIGM}        \nfig   
\fignam{\FIGA}        \nfig   
\fignam{\FIGB}        \nfig   
\fignam{\GRADA}       \nfig   
\fignam{\GRADB}       \nfig   
\fignam{\OH}          \nfig   
\def\infig#1#2#3{\epsfxsize=#3cm \centering{\mbox{\epsfbox{#2}}}}

\newcount\tabnumber\tabnumber=1
\def\ntab{\global\advance\tabnumber by 1}
\def\tabnam#1{\xdef#1{\the\tabnumber}}
\tabnam{\TDATA}        \ntab 
\tabnam{\TBAR}        \ntab 
\tabnam{\TSTAT}      \ntab 
\tabnam{\TMAG}      \ntab 
\newcount\eqtnumber\eqtnumber=1
\def\eqtnum{{\the\eqtnumber}\global\advance\eqtnumber by 1}
\newcount\neqtnumber\neqtnumber=1
\def\neqt{\global\advance\neqtnumber by 1}
\def\eqtnam#1{\xdef#1{\the\neqtnumber}}
\begin{document}

\thesaurus{03(11.11.1; 11.19.2; 11.19.3; 11.19.6)} 
\title{Starbursts in barred spiral galaxies}

\subtitle{V. Morphological analysis of bars
\thanks{Based on observations obtained at the 2 meter telescope of 
Observatoire du Pic du Midi, operated by INSU (CNRS)}
}

\author{S. Chapelon \inst{1,2} \and 
        T. Contini \inst{1,3}  \and 
        E. Davoust \inst{1}  
           }

\institute{
Observatoire Midi-Pyr\'en\'ees, UMR 5572, 14 Avenue E. Belin, F-31400 Toulouse, France
 \and
Laboratoire d'Astronomie Spatiale du CNRS, B.P. 8, 13376 Marseille Cedex 12, France
 \and
European Southern Observatory,
Karl-Schwarzschild-Str. 2, D-85748 Garching bei M\"unchen, Germany
             }

\offprints{Davoust, davoust@obs-mip.fr}

\date{Received ??; accepted ??}

\maketitle



\begin{abstract}
We have measured the bar lengths and widths of 125 barred galaxies observed 
with CCDs.  The dependence of bar strength (identified with bar axis ratio) on 
morphological type, nuclear activity, central and mid-bar surface brightness 
is investigated.  

The properties of the bars are best explained if the sample is divided into 
early- ($<$ SBbc) and late-type galaxies, and into active (starburst, Seyfert 
or LINER) and normal galaxies.  We find that galaxies with very long bars are 
mostly active and that normal late-type galaxies have a distinct behavior from 
the three other groups of galaxies.  We confirm earlier findings that active 
late-type galaxies tend to have both stronger and longer bars than normal 
ones.  An important result of this paper is that early-type galaxies
do not share this behavior : they all tend to have strong bars, whether
they are active or not. We also find 
correlations between bar strength and relative surface brightness in the 
middle and at the edge of the bar, which are not followed by normal late-type 
galaxies. 

These results are interpreted in the light of recent numerical simulations and 
paradigms about galaxy evolution.  They suggest that normal late-type galaxies 
represent the first stage of galaxy evolution, and that bars in early- and 
late-type galaxies do not have the same properties because they have a 
different origin. 

\end{abstract}  



\section{Introduction}

The presence of a bar strongly modifies the internal dynamics of spiral 
galaxies, as evidenced by the deviations from circular motion in the velocity 
maps of barred galaxies; the isovelocity curves in the bar tend to align with 
the bar major axis. Numerical simulations have shown that these non-circular 
motions induced by the presence of a bar are toward the center of the galaxy.  
Inside corotation, the dissipative gaseous component loses angular momentum to 
the stars and falls inward (e.g. Schwarz 1984, Wada \& Habe 1992, Friedli \& 
Benz 1993), as a result of the torque exerted by the bar. 

The rate of gaseous inflow depends essentially on the bar strength, which is 
the ratio of the force of the bar to that of the axisymmetric disk.  The bar 
strength in turn depends on the bar axis ratio and on its mass. The rate of 
inflow also depends on the bulge-to-disk mass ratio, but not on the bar 
pattern speed nor on the gas mass (Friedli \& Benz 1993). 

This bar driven fueling provides an obvious mechanism for producing the 
activity that is often observed in the center of barred spiral galaxies (e.g. 
Contini et al. 1998). However, the role of the bar on starburst (Hawarden et 
al. 1996) or Seyfert activity (Ho et al. 1997) in spiral galaxies remains 
controversial. 

In view of these conflicting results, a detailed study of the correlation 
between bar strength and starburst activity would be welcome.  A first step in 
this direction is the morphological study of Martin (1995), who measured the 
bar axis ratio of 136 barred galaxies on prints of blue photographs displayed 
in the atlas of Sandage \& Bedke (1988).  He found an apparent correlation 
between bar axis ratio and star formation activity.  Martinet \& Friedli 
(1997) selected 32 late-type galaxies from Martin's sample and showed that the 
galaxies with the strongest star formation activity had both thin and long 
bars, but that this was not a sufficient condition for violent star formation.  

The present paper is a further step in the study of the relation between bar 
strength and starbursts.  It is based on a morphological study of bars, and 
uses the bar axis ratio as an estimate of the bar strength, as in Martin (1995) 
and in Martinet \& Friedli (1997).  But several significant improvements are 
made.  Our investigation rests on CCD images, which lend themselves to more 
quantitative mesurements than photographic prints, and provide the photometric 
properties of the bars.  The choice of a red filter for the observations 
reduces the perturbing effect of dust. We implement a method for measuring the 
axis ratio which is not influenced by the perturbing presence of the bulge. 

Our main goal is to check whether the results of Martin (1995) and of 
Martinet \& 
Friedli (1997) still hold using our CCD data and method of measurement. Since 
our sample of galaxies is very different, biased toward starbursting and 
early-type galaxies, we can also hope to discriminate universal properties 
from those which are merely caused by selection effects.  Finally, we 
take advantage of the large database accumulated on our sample (Contini 1996, 
Contini et al. 1997) to look for correlations between bar morphology and other 
galaxian properties, such as starburst age, oxygen abundance, and neutral 
hydrogen mass. 

\section{The samples of barred galaxies} 

We used two samples of barred galaxies for this study. The first one is 
composed of all the Markarian galaxies that are barred and have been detected 
by IRAS. It contains 144 galaxies, for many of which we obtained CCD images, 
low-resolution CCD spectra, neutral hydrogen and CO profiles (Contini 1996).  

CCD images of 121 galaxies of this sample were obtained during several runs at 
the Bernard Lyot 2-meter telescope of Observatoire du Pic du Midi, with a 
1000$\times$1000 Thomson CCD (pixel size 0.24\arcsec\ on the sky).  The seeing 
was good, with a median value of about 1.5\arcsec. During one of those runs, 
we also observed NGC 6764 for one of the papers in this series (Contini et al. 
1997).  The details of the observations can be found in Contini (1996). 

From this sample, we selected the 100 galaxies which have a measurable bar. We 
decided to make the measurements on the images taken in the red (R' Cousins; 
for a definition, see de Vaucouleurs \& Longo 1988) band, because dust lanes 
which perturb the measurements are less conspicuous in that band than in the 
bluer ones. The calibration of the zero point of the magnitude scale was done 
by indirect procedures. The photometric constants of 34 galaxy images were 
obtained using published aperture photometry (de Vaucouleurs \& Longo 1988); 
the accuracy of the zero point is 0.1~mag. For the 66 other galaxies, we had 
to rely on a mean photometric constant for the night, with a resulting 
accuracy of only 0.4~mag.  

We also needed a comparison sample, with no bar or no starburst.
Since we did not have images of ordinary Markarian starburst galaxies, 
we took the second option and selected our comparison sample 
among the 113 galaxies observed by Frei et al. (1996), whose calibrated CCD 
images are publicly available bu ftp at astro.princeton.edu/frei/Galaxies.  We 
selected the galaxies of that sample that were classified barred in LEDA. 
After eliminating galaxies with no measurable bar (because of 
misclassification or high inclination), we were left with 26 galaxies.  We 
made the measurements on the images taken in the red (Cousins or Gunn) band, 
as for the first sample.  The zero point calibration of these images is
very good, with an average uncertainty of about 0.05 mag (if NGC 4498 is
excluded, which has an uncertainty of 0.4 mag.).

Table 1 gathers all the information on these galaxies necessary for our 
analysis. The galaxy name is in col. 1; the numerical morphological type and 
inclination (from LEDA) in col. 2 and 3; the distance (in Mpc, estimated from 
the redshift given in LEDA and a Hubble constant of 75 km s$^{-1}$ Mpc$^{-1}$) 
in col. 4; the absolute magnitude (from LEDA) in col. 5; the HI mass and FIR 
and H${\alpha}$ luminosities (in log of solar masses, solar luminosities and 
erg cm$^{-2}$s$^{-1}$ respectively, from Contini 1996) in col. 6, 7 and 8; the 
central (nearly central for Seyferts) oxygen abundance (from Contini et al. 
1998; from the literature for the comparison sample) in col. 9; the IRAS flux 
ratio $S_{\rm 25}/S_{\rm 100}$ (in log, computed from the fluxes given in 
Bicay et al. 1995, 
or the Faint Source Catalogue) followed by L (when an upper limit) in col. 10; 
the spectral classification (from Contini et al. 1998; from the literature for 
the comparison sample) in col. 11. No IRAS flux ratio is given when the fluxes 
at 25 and 100$\mu$m are both upper limits. Table 1 is given in electronic form 
only. 

\begin{table}
\caption[~\TDATA]{Physical parameters of the sample galaxies.
This Table is available in electronic form only} 
\end{table} 

Early-type galaxies are defined as earlier than SBbc (t$<$4), and late-types 
as SBbc and later.  The IRAS flux ratio is used to distinguish between 
starburst/Seyfert and more quiescent galaxies.  The former, which we hereafter 
call 'active',  have a ratio log($S_{\rm 25}/S_{\rm 100})  \ge$ -- 1.2; 
the others are called 'normal'. 
It is true that we have independently determined the spectral type 
of the Mrk galaxies by optical spectroscopy (Contini et al. 1998), but we have 
not done so for the sample of Frei et al. (1996), and we need a uniform 
criterion for both samples.  The adopted criterion also has the advantage of 
being the same as that used by Martinet \& Friedli (1997), and thus allows a 
comparison of our results with theirs.  This criterion is essentially a 
measure of the star formation efficiency.  Coziol et al. (1998) have in fact 
shown that IRAS colors are very efficient for separating starburst, quiescent 
and Seyfert galaxies. 

\section{Method of measurement of the bar parameters}

The method adopted by Martin (1995) for measuring the length and width of bars 
is visual and relies on photographic prints in the blue band. He estimates 
that, with this method, the uncertainty is about 20\%.  He defines the 
semi-major axis of the bar as the length from the galaxy center to the sharp 
outer tip where spiral arms begin, and the semi-minor axis as the length from 
the center to the edge of the bar, oval, lens or spheroidal component, 
measured perpendicularly to the major axis. 

Our CCD images allow us in principle to make a sophisticated numerical 
analysis, including adjustment of ellipses to the bar.  However, such a method 
is fraught with pitfalls which have been outlined by Wozniak et al (1995), and 
we preferred to rely on a simpler method. We first extract a 3-pixel wide 
photometric profile along the major axis of the bar.  The semi-major axis $a$ 
of the bar is the distance from the center of the galaxy to where the bar 
obviously ends. This is where the surface brightness profile abruptly changes 
slope to become steeper; this also coincides with the origin of the spiral 
arms. This measurement is not automatic, it relies on the subjective judgement 
of where the bar ends, and is comparable to that of Martin (1995). It is 
hopefully more reliable, since our CCD images have a large dynamic range and 
were taken in the red band. 

On the other hand, the method for measuring the width of the bar departs 
significantly from that of Martin (1995). We believe that his method 
overestimates the bar width, because it includes the effect of the bulge. To 
measure the width, we extract two photometric profiles (also three pixels 
wide) along the minor axis of the bar at two symmetrical distances of $a$/2 
from the center and take their average. The semi-minor axis $b$ of the bar is 
the distance from the bar major axis to the same isophotal level as where the 
tip of the bar was measured.  This method is comparable to that of Friedli \& 
Benz (1993). 

\section{Results of the bar measurements} 

The results of our bar measurements on CCD images are summarized in Table 2.  
The galaxy name is in col. 1; the position angle of its major axis (from LEDA; 
from our images if not in LEDA) in col. 2; the angle $\Phi$ between the bar 
and galaxy major axes (from our images) in col. 3; the measured semi-major and 
semi-minor axes (in arcsec) in col. 4 and 5; the deprojected semi-major and 
semi-minor axes (in arcsec) in col. 6 and 7; the deprojected axis ratio in 
col. 8; the ratio of deprojected major axis to corrected blue
diameter at the isophote 25 
($2a/D_c$, hereafter normalized bar length, where
$D_c$ is from LEDA) in col. 9; the surface 
brightness along the bar at the center ($\mu_c$), where the semi-minor axis is 
measured (at half the distance to the end of the bar, $\mu_m$) and at the tip 
of the bar ($\mu_b$), in mag.arcsec$^{-2}$ in col. 10, 11 and 12. Table 2 is 
given in electronic form only.  To give the reader a feeling for what
the different types of bars look like, we show examples of each category
in Fig.~\FIGM.

\begin{table}
\caption[~\TBAR]{Bar characteristics of the sample galaxies. 
This Table is available in electronic form only} 
\end{table} 

\begin{figure*}
\infig{16.16}{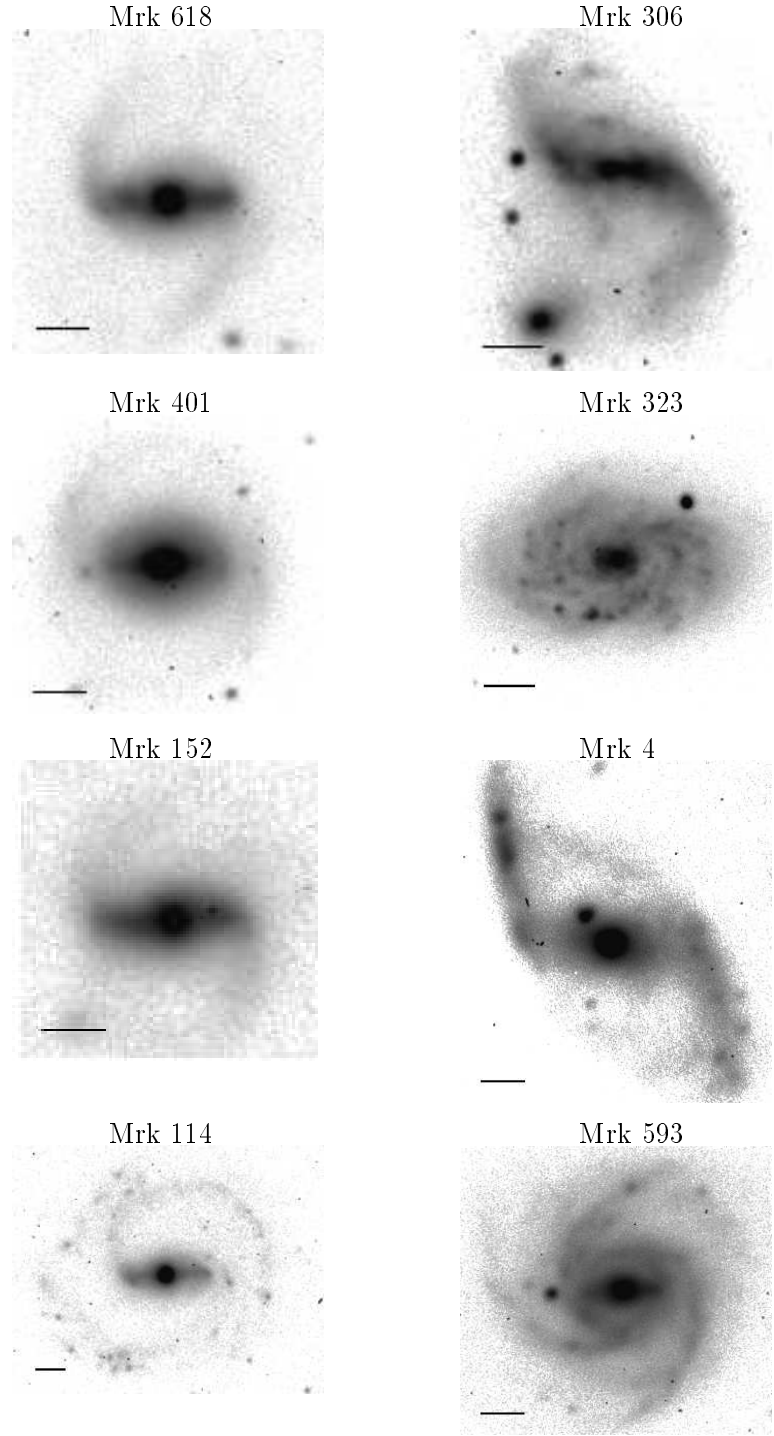}{16.16}
\caption[\FIGM]{Examples of each category of bars.  Early-type
galaxies are on the left and late-type ones on the right. From top to
bottom : strong, weak, long and short bars.  All images
have been flipped and rotated in such a way that the bar is horizontal
and the spiral pattern is winding counterclockwise.  The scale is
indicated by a 10-arcsec long horizontal line
}   
\end{figure*} 

The bar measurements were deprojected using the equations given by Martin 
(1995).  NGC 6764, Mrk 1326 and Mrk 1452 were observed in V only. For NGC 
6764, the bar surface brightnesses were transformed to R' using the mean 
(V -- R') color of the galaxy computed from de Vaucouleurs \& Longo (1988). 
For Mrk 1326 and 1452, the bar surface brightnesses were transformed to R' 
using the mean (V -- R')$_e$ for their respective morphological type (Buta \& 
Williams 1995). 

The fact that we chose to normalize the bar length by $D_c$ deserves an
explanation.  Martin (1995) and Martinet \& Friedli (1997) used
$D_{25}$, probably because they thought that absorption affects the
galaxy diameter and bar length in the same way. Our images, however, are
in the red bandpass, and, in all logic, we should use a red isophotal 
diameter.  We did measure the isophotal diameter $D_{24R}$ corresponding
to a surface brightness of 24 mag.arcsec$^{-2}$ on our images, and
compared it to $D_{25}$.  For the Frei sample, there is a good
correlation with a small scatter, from which we derived the mean
relation $D_{25} = 1.04\times D_{24R} - 0.03\times t$.
For the Mrk sample, on the other
hand, the scatter is much larger, because two thirds of the images
suffer from poor photometric calibration.  We thus had to resort to
$D_c$.  From the above relation, we estimate that the error induced by 
this choice (compared to a reliable $D_{24R}$) is not larger than 5\%, 
and reduced by the fact that we distinguish early- and
late-type galaxies in the analysis.

The repeatability of our measurements was involuntarily verified when we later 
discovered that NGC 4123, one of Frei et al.'s (1996) galaxies, is also Mrk 
1466. The comparison between the two sets of measurements for this galaxy 
gives an estimate of the external errors.  The relative uncertainty on the bar 
axis ratio is 20\%, while that on the bar length is less than 5\%. Since this 
galaxy is not in de Vaucouleurs \& Longo (1988), the errors on the estimated 
surface brightnesses reach 0.5 mag.  Only the measurements performed on our 
image of NGC 4123 were used in the subsequent analysis. 

We have compared our measurements with those of Kormendy (1979) for 13 
galaxies in common and those of Martin (1995) for 15 galaxies in common. We 
agree qualitatively with them, except for NGC 4487, where Martin (1995) finds 
a bar length and width half the size of ours; we checked that there was no 
scale error between the two images of that galaxy.  We find average 
differences ($a_{\rm us} - a_{\rm K})$ = --2.6\arcsec\ and 
($a_{\rm us} - a_{\rm M})$ = 2.8\arcsec. In other words, our estimates for the 
semi-major axis are intermediate between those of Martin and Kormendy.  These 
are marginally significant results because it is not obvious to decide exactly 
where the bar ends. Martin finds larger semi-minor axes than us by 2.0\arcsec\  
on the average, because he measures them from the center of the galaxy. For 
that reason, he also finds thicker bars than us by 0.14 on the average. 

The largest discrepancies with Martin (1995) are on the angle between the bar 
and galaxy major axes, which in turn produce large differences in the axis 
ratios corrected for inclination, $(b/a)_i$. 

We consider that $(b/a)_i$ is a good measure of the bar strength, and 
subsequently use the terms strong and weak rather than thin and thick. Strong 
bars are defined as having $(b/a)_i <$ 0.5; this is consistent with Martin 
(1995) who puts the limit at 0.6, but has weaker bars than us by 0.14 on the 
average. 

\section{Correlations with other galaxian properties} 

There are 125 different galaxies in the total sample, but in the astrophysical 
analysis of the data, we eliminated 23 galaxies for which the IRAS flux ratio 
log($S_{\rm 25}/S_{\rm 100}$) is either an upper limit (and much larger than 
-- 1.2) or unknown altogether.  In the analyses involving the deprojected bar 
axis ratio $(b/a)_i$, we also eliminated another 13 galaxies for which the 
deprojected bar axis ratio differed by more than 0.2 from the apparent one, 
because the uncertainties in the deprojection factor and in the true geometry 
of the bar mainly affect the deprojected semi-minor axis, and thus the 
deprojected bar axis ratio. 

\subsection{Bar characteristics and morphological type}

The mean value of the deprojected bar axis ratio for the total sample (125 
galaxies) is $(b/a)_i = 0.31 \pm 0.12$. For the sample restricted to reliable 
axis ratios and IRAS color (89 galaxies), it is 0.30. The 57 early-type 
galaxies seem to have a slightly stronger bar (0.28 $\pm$ 0.10) than the 32 
late-type ones (0.34 $\pm$ 0.14), but the difference is not statistically 
significant.  

The galaxies of our sample have on the average rather strong bars; less than 
5\% of them are in fact weak, compared to 36\% in Martin's (1995) sample 
(using the appropriate limit for each sample).  The 121 galaxies of his sample 
with a measured bar axis ratio have a mean ratio of 0.55 $\pm$ 0.19 (or about 
0.45 if we correct for the difference in method of measurement).  These 
differences in bar strength may be partly due to the fact that we have more 
early-type galaxies (60\% vs 18\%) and more active galaxies (57\% vs 36\%) 
than Martin (1995). The two samples are thus quite different. 

\begin{table}
\caption[~\TSTAT]{Mean values of the bar strength $(b/a)_i$ and
normalized length $2a/D_c$ for the different groups of galaxies of our
sample. $N$ is the number of galaxies in the group} 
\begin{flushleft}
\begin{tabular}{lrcccc}
\noalign{\smallskip}
\hline
\noalign{\smallskip}
Group&$N$&$(b/a)_i$&$N$&$2a/D_c$\\
\noalign{\smallskip}
\hline
\noalign{\smallskip}
t $\ge$ 4, normal&19&0.37 $\pm$ 0.18&23&0.22 $\pm$ 0.08\\
t $\ge$ 4, active&13&0.29 $\pm$ 0.08&14&0.35 $\pm$ 0.12\\
t $<$ 4, normal&19&0.29 $\pm$ 0.11&22&0.32 $\pm$ 0.14\\
t $<$ 4, active&38&0.27 $\pm$ 0.11&43&0.36 $\pm$ 0.14\\
\noalign{\smallskip}
\hline
\end{tabular}
\end{flushleft}
\end{table} 

There is marginal evidence that early-type galaxies have longer bars than 
late-type ones, relative to the galaxy size.  Such a difference is expected on 
dynamical grounds (Elmegreen \& Elmegreen 1985). The values of the normalized 
bar length $2a/D_c$ for the two types are 0.37 $\pm$ 0.17 and 0.28 $\pm$ 
0.12 respectively.  This confirms earlier findings by Elmegreen \& Elmegreen 
(1985) and Martin (1995). We have computed the mean bar lengths $2a/D_{25}$ of 
early- and late-type galaxies in Martin's sample, and find 0.33 and 0.17 
respectively. The late-type galaxies of Martin have much shorter bars than 
ours, but, since his galaxies are mostly normal, they should be compared to 
our normal late-type galaxies, which have a mean bar length of 0.23, thus 
quite near his value. 

\subsection{Bar characteristics and starburst activity}

The properties of the bars in relation to starburst activity are only revealed 
once the sample is broken into two groups, early- and late-type galaxies. 

Our results are summarized in Table 3, which gives the mean values of the bar 
strength and length for the different groups of galaxies. The numbers $N$ of 
galaxies in the different groups are not the same for the statistics on 
bar strength and length, because galaxies with large inclination correction 
were excluded from the analysis and statistics involving bar strength.  Our 
results are also shown in Figs. \FIGA\  and \FIGB, where the relation between 
bar strength and length for the two types of galaxies is plotted.  They reveal 
two main properties of bars with respect to activity: 

\begin{figure}
\infig{8.8}{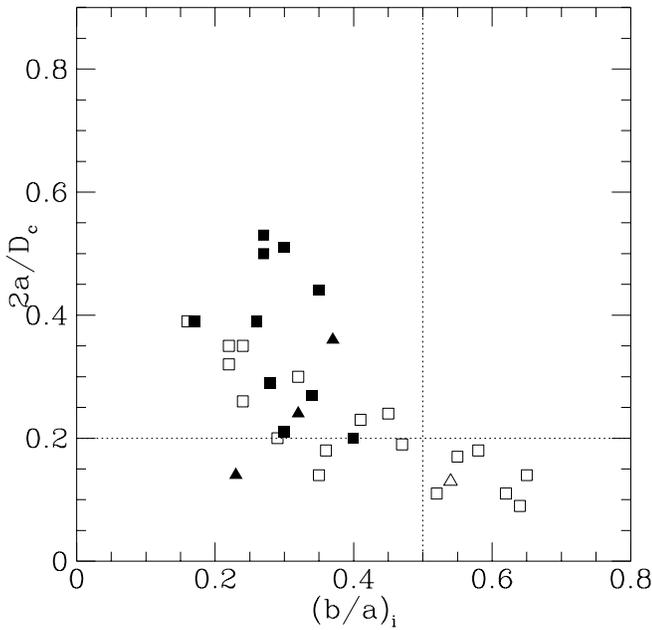}{8.8}
\caption[\FIGA]{Relation between bar strength and length in late-type
(t$\ge$4) galaxies. Filled symbols indicate galaxies with 
log($S_{\rm 25}/S_{\rm 100}) \ge$ -- 1.20 
and open symbols galaxies with log($S_{\rm 25}/S_{\rm 100}) <$ -- 1.20.  
Seyferts and LINERs are indicated by triangles, other galaxies by
squares. The limits between strong and weak bars and between long 
and short bars are indicated by dotted lines
}   
\end{figure} 

-- The normal late-type galaxies (empty squares in Fig.~\FIGA) stand out from 
the three other groups in that there are no very long bars ($2a/D_c > 0.4$).  
For this group, there is a good correlation between bar length and strength, 
of equation : $2a/D_c = -0.50(b/a)_i + 0.43$, with a
correlation coefficient of 0.85.

-- Galaxies with very long bars ($2a/D_c > 0.4$) are also strong and mostly 
active, while galaxies with short bars ($2a/D_c < 0.2$) are mostly normal. 

On the other hand, no trend of bar strength with activity appears, probably 
because our sample is biased toward strong bars. 

In Figs.~\FIGA\ and \FIGB, we distinguish Seyferts and LINERs (triangles) 
from starburst galaxies (squares). 
In fact, the two Figures show that the Seyferts and LINERs 
are undistinguishable from starbursts as far as bar 
characteristics are concerned. The only remarkable difference between the two 
types of active galaxies is that most of those with 
log($S_{\rm 25}/S_{\rm 100}) >$ -- 0.9 are Seyferts or LINERs.
It turns out that, with the adopted activity 
criterion, a small minority of Seyferts and LINERs falls in the category of 
normal galaxies.  

\begin{figure}
\infig{8.8}{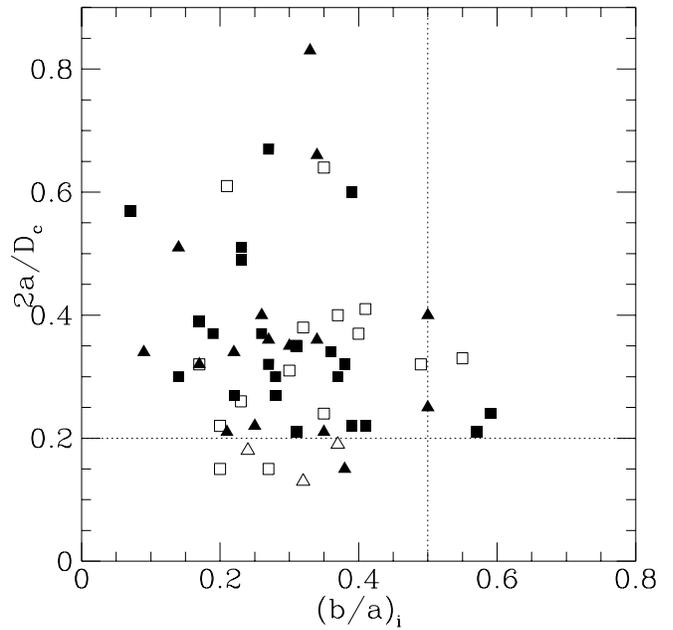}{8.8}
\caption[\FIGB]{Same as Fig.~\FIGA\ for early-type (t$<$4) galaxies
}   
\end{figure} 

We now turn to a comparison with results from the literature.  Using 32 
late-type galaxies from the sample of Martin (1995), Martinet \& Friedli 
(1997) showed that all the late-type galaxies with large star formation rates 
have both strong and long bars, but that some more quiescent galaxies also do 
share these properties. We confirm this result, using the 32 late-type 
galaxies with reliable $(b/a)_i$ of our sample, 5 of which are in common with 
Martinet \& Friedli's sample, but with independent measurements (see 
Fig.~\FIGA). On the other hand, this is not true for the 57 early-type 
galaxies of our sample (see Fig.~\FIGB).  Most of them, whether they are 
active or not, have strong and long bars, just like the active late-type 
galaxies. 

Our Fig. \FIGA\ should be compared to Fig. 3 of Martinet \& Friedli (1997).  
These authors chose a value of 0.18 for the limit between long and short bars; 
we adopted a value of 0.2 for comparison, since our bar estimates are slightly 
larger than those of Martin (1995) for galaxies in common.  Our Figs.~\FIGA, 
\FIGB\ and \OH\ suggest that a value of 0.4 would in fact be a more realistic 
limit for general purposes. 

The cumulative distribution of bar axis ratios shows that all the active 
late-type galaxies of our sample and 63 ($\pm$ 5)\% of the normal ones have 
strong bars.  Martin (1995) finds that about 71\% of his active galaxies and 
59\% of the quiescent ones have a strong bar; our active galaxies thus have 
much stronger bars than Martin's, but the strength of the quiescent ones are 
comparable. 

\subsection{Photometric properties of the bars}
We have measured the surface brightness along the bar at the center and at the 
distances $a$/2 and $a$ from the center. Hereafter we call these quantities  
$\mu_c$, $\mu_m$ and $\mu_b$ respectively; we also determined
$\Delta\mu_c =\mu_m-\mu_c$ and $\Delta\mu_b =\mu_b-\mu_m$. 
The reason for computing two surface brightness differences rather than
one is that the influence of the bulge is (hopefully) confined to the
first one only.  While $\mu_c$ 
depends on seeing, it is still a good indicator of the nuclear surface 
brightness.  We again break the sample into early- and late-type galaxies for 
comparing the photometric and morphological properties of the bars. The three 
main results are: 

\begin{figure}
\infig{8.8}{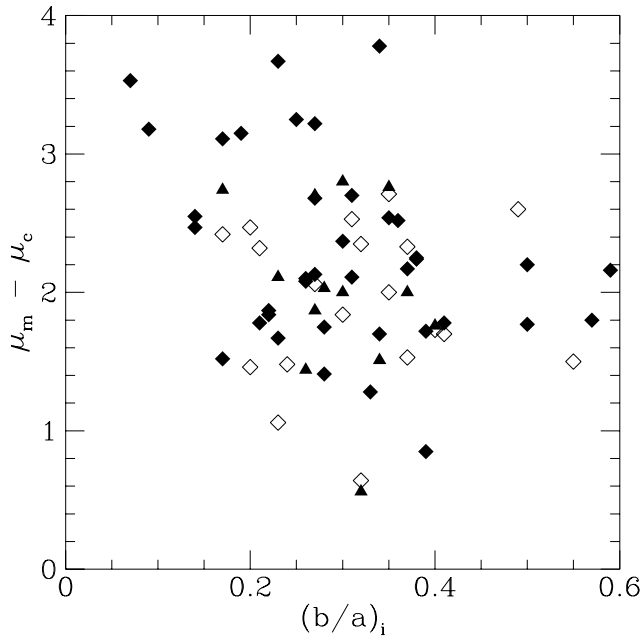}{8.8}
\caption[\GRADA]{Bar strength as a function of surface brightness 
difference between center and mid-bar, for early- (t$<$4; diamonds)
and late-type (triangles) galaxies; open symbols indicate normal galaxies
and filled ones active galaxies.  The normal late-type galaxies
have not been plotted because they do not share the correlation
}   
\end{figure} 

-- The normal late-type galaxies again stand out from the three other groups.  
They have a fainter central surface brightness ($\mu_c$) and a smaller 
$\Delta\mu_c$.  This is partly due to the fact that normal late-type galaxies 
have fainter $\mu_c$ and shorter bars than the other groups. The main result 
to keep in mind here is that the middle of the bar is about two magnitudes 
fainter than the nucleus, or about one magnitude fainter in the case of normal 
late-type galaxies. 

-- The surface brightness difference between middle and end of the bar 
($\Delta\mu_b$) is independent of morphological type or activity and is of 
about 0.44 mag. One can also note that $\Delta\mu_b$ is smaller than 
$\Delta\mu_c$ and that $\Delta\mu_c$ depends on morphological type for normal 
galaxies only (it is larger for early types); these last two results are due 
to the influence of the bulge which is more important in early-type galaxies.  

-- There is a correlation between bar strength and $\Delta\mu_b$, and an 
opposite one between bar strength and $\Delta\mu_c$, which are again shared by 
all groups except the normal late-types (see Figs.~\GRADA\ and \GRADB). 
Stronger bars have a larger $\Delta\mu_c$ and a smaller $\Delta\mu_b$, and 
vice versa. In other words, weak bars tend to have a flatter overall surface 
brightness gradient than strong ones.  

\begin{figure}
\infig{8.8}{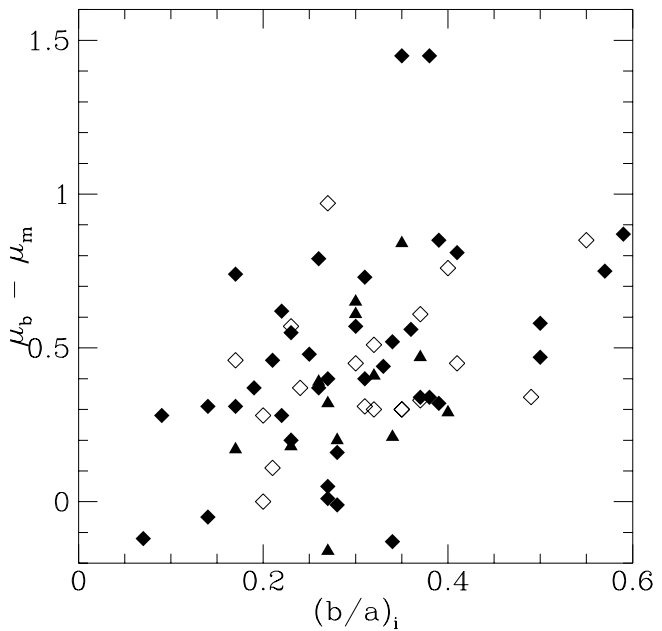}{8.8}
\caption[\GRADB]{ Same as Fig.~\GRADA\ for the surface
brightness difference between middle and end of bar. 
}   
\end{figure} 

The photometric properties of bars have already been studied in the past.  
Elmegreen \& Elmegreen (1985) and Elmegreen et al. (1996) find that bars in 
early-type galaxies tend to have flat\footnote{Note that flat does
not mean constant but straight} light profiles and those in late-type 
galaxies tend to have exponential profiles outside the bulge region.  
Comparisons with our results are difficult because the nature of these surface 
brightness gradients is determined by comparison with the gradient in the 
disk, and the bar strength is not determined.  Nevertheless, using bar 
strength determined by Martin (1995) for the 4 late-type galaxies with 
exponential profiles (NGC 3359, which has a very strong bar, is considered to 
have a flat profile in their second paper), we find a mean value of 0.42, thus 
rather weak, which is expected for bars with steep outer brightness profiles.  
Ohta et al. (1990) have determined the surface brightness profiles of 6 
early-type galaxies.  They find rather flat outer gradients, less than 0.5 
mag, for 4 out of 6 galaxies; all of them have strong bars. 

\subsection{Bar strength and central oxygen abundance}

We have investigated a possible dependence of bar characteristics on the 
oxygen abundance measured in the center of the galaxies. The relation between 
bar length and central oxygen abundance (in solar units)
for early- and late-type galaxies is 
displayed in Fig.~\OH.  While there seems to be no trend with bar strength, we 
find that there are no galaxies with high oxygen abundance (O/H $>$ 1.4) and 
very long ($2a/D_c > 0.4$) bars. 

\begin{table}
\caption[~\TMAG]{Mean values of the  nuclear surface brightness
($\mu_c$) and of the surface brightness differences 
between center and mid-bar ($\Delta\mu_c$) and between middle and end of
bar ($\Delta\mu_b$) for the
different groups of galaxies of our sample. $N$ is the number of galaxies 
in the group} 
\begin{flushleft}
\begin{tabular}{lrcccc}
\noalign{\smallskip}
\hline
\noalign{\smallskip}
Group&$N$&$\mu_c$&$\Delta\mu_c$&$\Delta\mu_b$\\
\noalign{\smallskip}
\hline
\noalign{\smallskip}
t $\ge$ 4, nor&23&18.68 $\pm$ 0.83&0.95 $\pm$ 0.41&0.37 $\pm$ 0.20\\
t $\ge$ 4, act&14&17.86 $\pm$ 0.94&2.00 $\pm$ 0.62&0.42 $\pm$ 0.34\\
t $<$ 4, nor&22&17.96 $\pm$ 0.63&1.94 $\pm$ 0.53&0.48 $\pm$ 0.25\\
t $<$ 4, act&43&17.85 $\pm$ 0.89&2.25 $\pm$ 0.70&0.46 $\pm$ 0.34\\
\noalign{\smallskip}
\hline
\end{tabular}
\end{flushleft}
\end{table} 

We have also studied the relation between bar length and strength and the 
oxygen abundance gradient along the bar. The results are described in paper IV 
of this series (Consid\`ere et al., in preparation). 

A search for correlations between bar characteristics and blue, far infrared 
and H$\alpha$ luminosities and neutral hydrogen mass did not reveal any 
meaningful trends, even when normalizing these quantities by the blue 
luminosity. 

\section{The role of bars in the evolution of galaxies}

We now use the bar properties determined in the previous section to discuss 
the possible role of bars in shaping the evolution of galaxies. 

The first point is to understand why normal late-type galaxies stand out from 
the three other groups.  This can be seen in two ways, depending on whether 
the activity or the morphology of the galaxy is considered the determining 
factor.  In the first case, the normal early-type galaxies stand out because 
they behave like active galaxies; in the second case, the active late-type 
galaxies should be singled out, because they have the same properties as the 
early-type galaxies. In either case one reaches the conclusion that early- and 
late-type galaxies differ in their bar properties, and thus presumably in the 
way the bar originates and the galaxy evolves. 

This concurs with the results of numerical simulations by Noguchi (1996), who 
concludes that the bars of late-type galaxies slowly form as a result of bar 
instability in the disk, whereas the bars of early-type galaxies form more 
rapidly (within one disk rotation period) in tidal interactions.  He also 
finds that tidally induced bars have flatter surface density profiles than 
spontaneous bars. Even though he does not study the dependence of this profile 
on bar strength, there are indications in his paper that induced bars are 
stronger than spontaneous bars, which is expected if interactions induce a 
stronger perturbation than natural disk instabilities.  This would confirm the 
relation that we find between strength of the bar and flatness of the outer 
surface brightness profile. 

\begin{figure}
\infig{8.8}{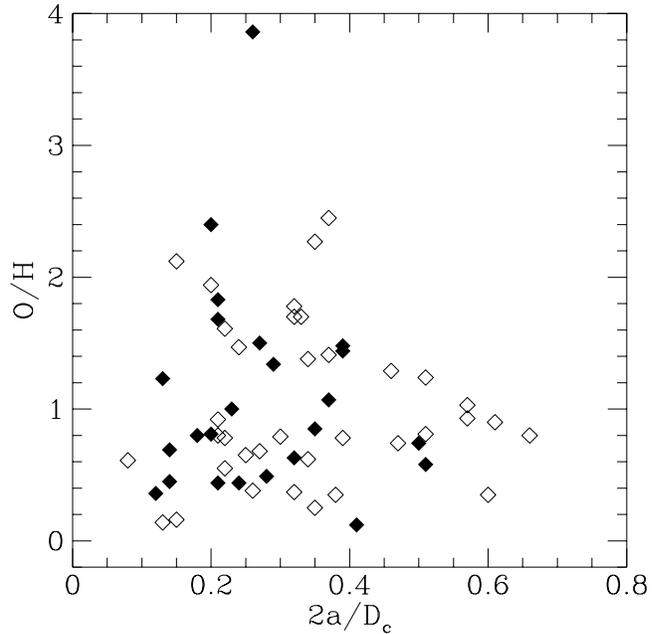}{8.8}
\caption[\OH]{Normalized bar length as a function of central oxygen abundance 
(in solar units) for early- (open diamonds) and late-type (filled diamonds) 
galaxies.  There are no galaxies with long bars and high oxygen abundance
}   
\end{figure} 

Now, to explain the specificity of normal late-type galaxies, one has to 
consider the overall evolution of galaxies.  According to the latest paradigm 
(Pfenniger 1998), galaxies evolve from late to early type by forming a bar 
which dissolves into a bulge, then grows again to form a more massive bulge.  
In this scenario, a normal late-type galaxy, which initially develops a short 
bar, becomes active as the bar gets stronger and longer and funnels more gas 
toward the center to feed star formation.  In the later evolution, a new and 
initially strong bar arises in the wake of a tidal perturbation, grows longer 
and star formation resumes after sufficient gas has been accreted. The 
evolution thus proceeds, alternating between barred and unbarred structure, 
until there is no more gas to be funnelled toward the center. 

The correlation between bar strength and $\Delta\mu_c$ is readily understood 
in this context: a stronger bar favors central star formation, which in turn 
enhances the brightness contrast between the bar and the nucleus. The fact 
that stronger bars have flatter outer profiles, predicted by the simulations, 
is probably best understood in terms of stellar dynamics.  In strong bars, the 
stellar orbits are more elongated and the stars spend more time near 
apogalacticon, thus enhancing the surface brightness of the outer regions of 
the bar. However, this explanation rests on the assumption that these are 
regular orbits; chaotic orbits are likely to fill larger regions and to smooth 
out the density contrast (Wozniak \& Pfenniger 1999). 

Finally, the absence of very long bars with high oxygen abundances can be 
explained in two different ways.  The first one, which was suggested to us by 
Daniel Friedli, is that long bars allow gas from regions further out, where 
the oxygen abundances are very low, to reach the central regions and thus to 
dilute the central abundances. Alternatively, star formation increases the 
amount of dust, which can hide the extremities of the bars. We have indeed 
noticed that the bars of our galaxies are longer in K than in the R band 
(Bergougnan et al., in preparation); a similar trend has also been noticed by 
Friedli et al. (1996). We are indebted to Herv\'e Wozniak for pointing this 
out to us.  The correct explanation might be a combination of these two 
possibilities. 

\begin{acknowledgements}
Data from the literature were obtained with the Lyon Meudon Extragalactic 
database (LEDA), supplied by the LEDA team at CRAL-Observatoire de Lyon 
(France).  We thank the staff of Observatoire du Pic du Midi 
for assistance at the telescope. We also thank Christophe Bordry and David 
Teyssier for their contribution to the statistical analysis of the data,
as well as Alessandro Boselli, Daniel Friedli, Herv\'e Wozniak and Ron Buta
for helpful comments on this paper. 
 \end{acknowledgements}


\begin{thebibliography}{}

\bibitem{} 
\bibitem{} Bicay M.D., Kojoian G., Seal J., et al., 1995, ApJ 98, 369
\bibitem{} Buta R., Williams K.L., 1995, AJ 109, 543
\bibitem{} Contini T., 1996, Ph. D. thesis, Universit\'e Paul Sabatier, 
Toulouse, France
\bibitem{} Contini T., Consid\`ere S., Davoust E., 1998, A\&AS 130, 285 
(paper III)
\bibitem{} Contini T., Davoust E., Consid\`ere S., 1995, A\&A 303, 440 (paper I)
\bibitem{} Contini T., Wozniak H., Consid\`ere S., Davoust E., 1997, A\&A, 
324, 41 (paper II)
\bibitem{} Coziol R., Contini T., Davoust E., Consid\`ere S., 1997, ApJ,
481, L67
\bibitem{} Coziol R., Torres C.A.O, Quast G.R., Contini T., Davoust E., 1998, 
ApJS, 119, 239
\bibitem{} Elmegreen B.G., Elmegreen D.M., 1985, ApJ 288, 438
\bibitem{} Elmegreen B.G., Elmegreen D.M., Chromey F.R., et al., 1996, AJ 
111, 2233
\bibitem{} Frei Z., Guhathakurta P., Gunn J., Tyson J.A., 1996, AJ 111, 174
\bibitem{} Friedli D., Benz W., 1993 A\&A 268, 65
\bibitem{} Friedli D., Wozniak H., Rieke M., et al., 1996, A\&AS 118, 461
\bibitem{} Hawarden T.G., Huang J.H., Gu Q.S., 1996, in : Barred
Galaxies, Buta R., Crocker D.A., Elmegreen B.G. (eds.), Proc. IAU Coll.
157, ASP Conference Series, p. 54
\bibitem{} Ho L.C., Filippenko A.V., Sargent W.L.W., 1997, ApJ 487, 591
\bibitem{} Kormendy, J., 1979, ApJ 277, 714
\bibitem{} de Vaucouleurs A., Longo G., 1988, The University of Texas 
Monographs in Astronomy No. 5
\bibitem{} Martin P., 1995, AJ 109, 2428
\bibitem{} Martinet L., Friedli D., 1997, A\&A, 323, 363
\bibitem{} Noguchi M., 1996, ApJ 469, 605
\bibitem{} Ohta K., Hamabe M., Wakamatsu K.I., 1990, ApJ 357, 71
\bibitem{} Pfenniger D., 1998, in : Abundance profiles: diagnostics tools
for galaxy history, Friedli D., Edmunds M.G., Robert C., Drissen L. (eds.), 
ASP Conference Series, p. 237
\bibitem{} Sandage A., Bedke J., 1988, Atlas of galaxies useful for
measuring the cosmological distance scale (NASA, Washington DC)
\bibitem{} Schwarz, M.P., 1984, MNRAS 209, 93
\bibitem{} Wada K., Habe A., 1992, MNRAS 258, 82
\bibitem{} Wozniak H., Friedli D., Martinet L., Martin P., Bratschi P.,
1995, A\&AS 111, 115
\bibitem{} Wozniak H., Pfenniger, D., 
1999, in:"Impact of Modern Dynamics in Astronomy", Proc IAU Coll. 172,
J. Henrard \& S. Ferraz-Mello (eds), Celestial Mechanics, in press
\end{thebibliography}
\end{document}


{\bf Table 1:} Physical characteristics of the sample galaxies

\begin{flushleft}
\begin{tabular}{lrrrrrrrrll}
\noalign{\smallskip}
\hline
\noalign{\smallskip}
Name&Type&Incl.&Dist.&M$_{abs}$&M(HI)&L(FIR)&L(H$\alpha$)&O/H&
log($S_{\rm 25}/S_{\rm 100})$&Sp.t\\
(1)&(2)&(3)&(4)&(5)&(6)&(7)&(8)&(9)&(10)&(11)\\
\noalign{\smallskip}
\hline
\noalign{\smallskip}
Mrk 1&  3.1&  55.1&  66.2& -19.77&  9.57& 10.13&&& -0.53  &Sy2\\
Mrk 2&  0.6&  31.7&  75.8& -21.00&  9.47& 10.64&& 1.61& -0.91  &STB\\
Mrk 4&  6.0&  62.4&  73.4& -20.02&  9.57& 10.37& 40.30&& -1.19  &STB\\
Mrk 10&  3.1&  70.2& 119.4& -22.96& 10.63& 10.34&& 0.65& -0.95  &Sy1\\
Mrk 12&  5.0&  38.0&  55.9& -21.24&  9.97& 10.37&& 0.80& -1.38  &STB\\
Mrk 13&  3.1&  32.5&  21.7& -17.84&  9.55&  8.59&  39.54& 0.61&&STB\\
Mrk 21&  3.6&  49.2& 115.1& -20.96&& 10.11&  39.59& 1.03& -1.08  &STB\\
Mrk 42&  2.7&  38.0&  98.5& -20.01&&  9.80&  40.40& 0.93&&Sy1\\
Mrk 52& -0.7&  61.6&  28.5& -19.40&  8.99&  9.70&  40.84& 0.78& -0.73  &STB\\
Mrk 79&  3.0&   3.8&  90.8& -21.65&  9.89& 10.24&&& -0.49  &Sy2\\
Mrk 87&  1.0&  45.3&  51.0& -20.22&  9.86&  9.79&  40.14&& -1.49  &STB\\
Mrk 90&  3.8&  29.6&  59.4& -20.06&  9.17&  9.74&  40.64&& -1.14  &STB\\
Mrk 109&  8.0&  49.9& 123.4& -20.14& 10.15& 10.20&0.42&& -0.80L &STB\\
Mrk 114&  3.1&  35.8& 103.1& -21.29& 10.30& 10.41&  40.94&& -1.28  &STB\\
Mrk 122&  2.0&  66.4&  90.4& -21.25&& 10.20&  40.25&& -1.40  &STB\\
Mrk 152&  2.7&  67.4&  94.3& -20.77&&  9.94&  40.70&& -0.73  &LIN\\
Mrk 161&  4.3&  52.3&  82.1& -21.19&  9.99& 10.49&  41.07& 0.74& -1.05  &STB\\
Mrk 179&  5.3&  39.7&  47.4& -20.07&  9.34&  9.59&  39.37& 1.44& -1.15  &STB\\
Mrk 185&  6.0&  42.6&  43.9& -20.80&  9.95&  9.91&  39.64&& -1.50  &STB\\
Mrk 188&  5.3&  39.5&  35.0& -20.48&  9.69& 10.00&  40.45& 3.86& -1.41  &STB\\
Mrk 213&  1.1&  49.2&  44.7& -20.61&  9.40& 10.07&  40.74& 0.37& -0.99  &STB\\
Mrk 271&  3.1&  43.9& 103.9& -21.89&& 10.49&  41.15& 0.14& -1.37  &Sy2\\
Mrk 281&  3.1&  36.3&  32.5& -20.85&  9.74&  9.99&&& -1.30  &STB\\
Mrk 291&  1.0&  49.1& 143.6& -20.57&& 10.16&&&&Sy1\\
Mrk 300& -0.1&  59.6& 157.5& -21.20&  9.85& 10.43&  42.14& 0.74&&STB\\
Mrk 306&  4.3&  52.3&  76.5& -20.69&  9.86&  9.87&  40.39& 0.63& -1.14L &STB\\
Mrk 307&  4.6&  37.2&  75.9& -21.33&  9.82& 10.26&& 1.00& -1.35  &STB\\
Mrk 319&  1.0&  55.8& 109.7& -21.63&  9.98& 10.99&&& -1.12  &STB\\
Mrk 323&  6.0&  43.8&  59.5& -20.64&  9.49& 10.30&  40.31&& -1.36  &STB\\
Mrk 326&  4.0&  49.8&  49.2& -20.19&  9.75& 10.13&& 1.83& -0.92  &STB\\
Mrk 332&  4.7&  24.8&  33.6& -20.03&  9.06&  9.95&  40.17& 0.81& -1.18  &STB\\
Mrk 339&  4.0&   6.3&  71.5& -20.01&  9.73&  9.75&&& -0.96  &LIN\\
Mrk 353&  3.4&  58.5&  63.6& -20.35&  9.33& 10.36&  41.59& 0.55& -1.00  &STB\\
Mrk 358&  4.0&  42.7& 182.5& -21.84& 10.17& 10.44&&&&Sy1\\
Mrk 359&  3.4&  36.2&  68.8& -20.46&  8.85&  9.92&&& -0.60  &Sy1\\
Mrk 373&  4.0&  44.5&  81.2& -19.73&  9.27& 10.23&  41.58& 0.58& -0.93  &STB\\
Mrk 382&  4.6&  30.0& 136.7& -21.09&  9.77&  9.96&&&&Sy1\\
Mrk 384&  3.0&  50.6&  63.2& -20.79&  9.29& 10.37&  40.06&& -1.07  &STB\\
Mrk 386&  3.8&  66.8&  48.3& -20.96&  9.90&  9.51&&& -1.03L &STB\\
Mrk 401&  0.5&  27.4&  23.7& -18.31&  8.82&  9.30&  39.12& 1.47& -0.85  &STB\\
Mrk 446&  3.1&  40.6&  96.6& -21.42&  9.91& 10.31&  40.52& 0.80& -0.95  &STB\\
Mrk 474&  0.0&  49.8& 154.4& -20.73&& 10.33&&& -0.73L &Sy1\\
Mrk 479&  4.0&  60.6&  82.6& -21.18& 10.04& 10.27&  41.28& 0.45& -1.36  &STB\\
Mrk 493&  3.0&  14.3& 128.6& -20.97&  9.82& 10.23&&& -0.83  &Sy1\\
Mrk 533&  3.8&  24.5& 117.5& -21.81& 10.35& 11.04&&& -0.63  &Sy2\\
Mrk 538&  3.1&  41.7&  37.8& -20.38&  9.81& 10.30&&& -0.61  &LIN\\
Mrk 545&  1.1&  49.1&  62.6& -21.67&  9.90& 10.73&  41.65& 0.92& -1.15  &STB\\
Mrk 571&  3.4&  48.6&  68.6& -20.78&  9.80& 10.04&  41.33&& -1.07L &STB\\
Mrk 575&  1.0&  39.2&  73.7& -20.92&  9.69& 10.40&  40.41&& -1.04  &STB\\
Mrk 592&  3.0&  37.2& 100.9& -21.27&  9.92& 10.35&&& -0.93  &STB\\
Mrk 593&  5.3&   6.6& 110.5& -21.05& 10.22& 10.31&  40.54&& -0.79L &STB\\
Mrk 602&  4.0&  36.8&  37.6& -19.52&  9.71&  9.87&  41.38& 1.34& -0.95  &STB\\
\noalign{\smallskip}
\hline
\end{tabular}
\end{flushleft}

\begin{flushleft}
\begin{tabular}{lrrrrrrrrll}
\noalign{\smallskip}
\hline
\noalign{\smallskip}
(1)&(2)&(3)&(4)&(5)&(6)&(7)&(8)&(9)&(10)&(11)\\
\noalign{\smallskip}
\hline
\noalign{\smallskip}
Mrk 617&  4.9&  35.1&  62.3& -20.93&  9.53& 11.20&  42.18& 0.69& -0.65  &LIN\\
Mrk 618&  3.0&  42.8& 137.9& -22.18&  9.89& 10.87&& 0.81& -0.73  &Sy1\\
Mrk 620&  1.0&  44.3&  27.2& -20.23&  9.29&  9.83&  41.09& 0.16& -0.87  &Sy2\\
Mrk 665& -1.4&  32.1& 107.4& -21.07&  9.58& 10.16&  41.35& 0.68& -1.06  &STB\\
Mrk 686&  1.3&  47.6&  58.9& -20.72&  9.18&  9.63&& 0.12& -1.14  &Sy2\\
Mrk 691&  5.3&  53.4&  45.9& -21.02&  9.93& 10.12&  41.14& 1.50& -1.16  &STB\\
Mrk 703&  2.1&   2.1&  50.7& -20.74& 10.01& 10.19&&& -1.09  &STB\\
Mrk 710&  2.5&  51.0&  19.8& -18.71&  9.12&  9.18&  40.09& 1.24& -1.00  &STB\\
Mrk 712&  3.6&  62.8&  61.6& -20.64&  9.75&  9.70&  40.78& 0.35& -0.95  &STB\\
Mrk 731& -1.3&  49.0&  19.0& -18.55&  8.14&  9.19&  39.25& 1.94& -0.78L &STB\\
Mrk 752&  4.1&  35.3&  81.5& -20.83&  9.85&  9.90&  40.67& 0.44& -0.80L &STB\\
Mrk 759&  5.2&  37.7&  29.7& -20.13&  9.61&  9.79&  40.33&& -1.21  &STB\\
Mrk 781&  4.6&  31.5&  38.3& -20.11&  9.50&  9.66&  39.94& 0.49& -0.96L &STB\\
Mrk 799&  3.0&  51.4&  42.8& -20.91&  9.59& 10.48&& 2.27& -1.08  &STB\\
Mrk 814&  3.2&  70.6&  54.3& -20.43&  9.80&  9.83&  40.03& 0.38& -1.25  &STB\\
Mrk 861&  4.9&  21.3&  60.5& -19.87&  9.20&  9.84&&& -0.94  &STB\\
Mrk 871&  4.9&  60.3& 136.8& -21.74&  9.87& 10.30&&& -0.49  &Sy1\\
Mrk 874&  3.6&  59.5&  58.8& -19.79&  9.33&  9.35&  40.35& 0.35& -1.16  &STB\\
Mrk 896&  2.7&  56.0& 103.2& -21.70&& 10.05&&& -0.91  &Sy1\\
Mrk 898&  3.1&  52.3&  70.1& -20.08&  9.16&  9.79&& 1.78& -0.75L &STB\\
Mrk 917&  1.0&  21.6& 100.0& -20.98&  9.38& 10.74&&& -0.99  &Sy2\\
Mrk 955&  4.2&  57.3& 139.2& -21.47&  9.64& 10.49&&& -0.64L &Sy2\\
Mrk 968&  3.7&  23.3&  59.3& -20.52&  9.84&  9.66&&& -0.77L &STB\\
Mrk 1009&  3.1&  54.7&  57.1& -21.15&  9.89&  9.82&  40.35& 0.78& -1.29  &STB\\
Mrk 1050&  1.0&  54.2&  67.3& -20.61&  9.59& 10.52&  41.68&& -1.03  &STB\\
Mrk 1058&  4.8&  64.1&  70.2& -19.94&&  9.74&  41.51& 0.12& -0.91  &Sy2\\
Mrk 1066& -1.1&  54.7&  50.0& -20.23&& 10.54&  41.97& 0.25& -0.73  &Sy2\\
Mrk 1067&  1.6&  41.0&  57.7& -20.53&  9.67& 10.07&  40.42&& -1.43  &STB\\
Mrk 1073&  3.0&  21.5&  95.3& -21.56&  9.53& 11.02&  41.49& 0.62& -0.90  &Sy2\\
Mrk 1076&  3.9&  64.3&  95.8& -20.48&&  9.85&  41.45& 1.29&&STB\\
Mrk 1086&  2.7&  27.8& 118.9& -21.21&  9.92& 10.25&  40.05& 0.79& -0.98  &STB\\
Mrk 1088&  0.1&  20.7&  60.6& -21.82&  9.43& 10.53&  41.82& 1.38& -1.11  &STB\\
Mrk 1149&  0.0&  59.9&  83.5& -19.85&  9.58&  9.96&&&&STB\\
Mrk 1157& -0.1&  38.5&  62.7& -20.32&  9.38& 10.06&&& -0.89  &Sy2\\
Mrk 1180&  0.0&  43.7&  62.1& -21.04&&  9.90&&& -1.59L &STB\\
Mrk 1194& -2.0&  52.7&  59.0& -21.30&  9.47& 10.56&  41.80&& -1.21  &STB\\
Mrk 1200&  2.7&   7.9& 103.6& -20.96&  9.77& 10.38&  40.65&& -1.03  &LIN\\
Mrk 1231&  0.5&  52.6& 149.6& -21.14&& 10.60&  40.04&& -1.00  &STB\\
Mrk 1273&  0.0&  57.9& 108.0& -20.46&& 10.11&  40.13&& -0.91L &STB\\
Mrk 1291&  3.9&  36.8&  48.1& -20.84& 10.08&  9.89&&& -1.31  &Sy2\\
Mrk 1302&  3.0&  52.3&  74.5& -20.35&  9.77&  9.68&& 0.90& -0.52L &STB\\
Mrk 1326&  5.9&  27.2&  18.9& -18.58&  8.73&  8.83&  38.82& 1.68& -0.67L &STB\\
Mrk 1341&  5.9&  47.7&  15.7& -18.91&  9.03&  9.04&  39.02& 0.85& -1.30  &STB\\
Mrk 1433&  4.1&  39.2& 256.4& -22.59&& 10.95&  41.35&& -1.30  &STB\\
Mrk 1452&  4.9&  39.0&  92.4& -19.90&  9.70&  9.90&  39.31& 1.07&&STB\\
Mrk 1466&  5.1&  42.0&  17.7& -19.85&  9.65&  9.45&  40.41& 1.48& -0.93  &STB\\
Mrk 1485&  3.6&  38.5&  33.4& -20.82&  9.90&  9.74&  39.71& 2.12& -1.43  &STB\\
NGC 2541&  6.0&  64.7&   9.5& -18.77&  9.48&  9.18&& 0.36& -0.61L &   \\
NGC 2715&  5.2&  74.1&  21.0& -21.03&  9.91&  9.33&&& -1.80  &   \\
NGC 3319&  6.0&  57.6&  11.9& -19.52&  9.46&  8.46&& 0.44&&   \\
NGC 3351&  3.1&  55.9&  10.6& -20.30&  9.21&  9.49&& 2.45& -0.91  &STB\\
NGC 3368&  1.8&  51.1&  12.2& -20.96&  9.41&  9.42&& 1.41& -1.72  &   \\
NGC 3486&  5.2&  46.0&  10.4& -19.63&  9.56&  9.01&&& -1.68  &   \\
NGC 3726&  5.2&  48.2&  14.2& -20.49&  9.78&  9.20&&& -1.34  &   \\
NGC 3953&  4.0&  61.7&  16.8& -21.20&  9.33&  9.83&&& -1.40  &   \\
\noalign{\smallskip}
\hline
\end{tabular}
\end{flushleft}

\begin{flushleft}
\begin{tabular}{lrrrrrrrrll}
\noalign{\smallskip}
\hline
\noalign{\smallskip}
(1)&(2)&(3)&(4)&(5)&(6)&(7)&(8)&(9)&(10)&(11)\\
\noalign{\smallskip}
\hline
\noalign{\smallskip}
NGC 4088&  4.2&  72.0&  12.8& -20.43&  9.64&  9.73&&& -1.36  &   \\
NGC 4123&  5.1&  42.0&  17.7& -19.85&  9.65&  9.45&  40.41& 1.48& -0.93  &STB\\
NGC 4136&  5.2&  22.4&   9.7& -18.42&  9.04&  8.41&&& -1.26L &   \\
NGC 4178&  7.2&  72.0&   5.5& -19.70&  9.02&  8.46&&& -1.57L &   \\
NGC 4189&  5.9&  47.3&  28.8& -20.40&  9.51&  9.69&&& -1.35  &   \\
NGC 4303&  4.0&  25.4&  21.1& -21.88& 10.10& 10.28&& 1.23& -1.66  &LIN\\
NGC 4394&  2.9&  18.9&  13.3& -19.28&  8.47&  8.63&&& -1.44  &   \\
NGC 4487&  6.0&  46.6&  13.2& -19.50&  9.17&  8.85&&& -1.18L &   \\
NGC 4498&  6.3&  59.4&  21.0& -19.60&  9.11&  9.03&&& -1.35L &   \\
NGC 4548&  3.1&  36.8&   7.3& -20.99&  8.17&  8.31&&& -1.74L &   \\
NGC 4579&  2.8&  38.4&  20.9& -21.67&  8.99&  9.65&&& -1.53  &Sy1\\
NGC 4593&  3.0&  47.1&  32.8& -21.29&  9.33&  9.69&&& -0.82  &Sy1\\
NGC 4654&  5.9&  53.7&  14.6& -20.46&  9.46&  9.71&& 2.40& -1.35  &   \\
NGC 4725&  2.4&  48.9&  17.7& -21.77&  9.94&  9.53&& 1.70& -2.02  &   \\
NGC 4731&  5.9&  69.4&  19.5& -20.55& 10.03&  9.22&&& -1.56L &   \\
NGC 5377&  1.0&  60.0&  26.8& -20.38&  9.23&  9.14&&& -1.32L &   \\
NGC 5850&  3.1&  36.5&  34.8& -21.62&  9.76&  9.42&&& -1.54L &   \\
NGC 6384&  3.9&  51.0&  23.6& -21.72& 10.10&  9.53&& 1.70& -1.84  &   \\
NGC 6764&  3.6&  58.1&  35.7& -21.14&  9.67& 10.10&& 0.80& -0.95  &LIN\\
\noalign{\smallskip}
\hline
\end{tabular}
\end{flushleft}